\documentclass[aps,prl,twocolumn,superscriptaddress,showpacs]{revtex4}

\usepackage{graphicx}
\usepackage{amssymb}

\begin{document}

\title{Resonant light scattering by optical solitons}

\author{S. Flach}

\affiliation{Max-Planck-Institut f\"ur Physik komplexer Systeme,
N\"othnitzerstr. 38, Dresden 01187, Germany}

\author{V. Fleurov}

\affiliation{Beverly and Raymond Sackler Faculty of Exact Sciences,
School of Physics and Astronomy, Tel Aviv University, Tel Aviv 69978, Israel}

\author{A.V. Gorbach}

\affiliation{Max-Planck-Institut f\"ur Physik komplexer Systeme,
N\"othnitzerstr. 38, Dresden 01187, Germany}

\author{A.E. Miroshnichenko}

\affiliation{Nonlinear Physics Centre,
Research School of Physical Sciences and Engineering,
Australian National University, Canberra ACT 0200, Australia}

\begin{abstract}
We consider the process of light scattering by optical solitons in a planar waveguide
with homogeneous and inhomogeneous refractive index core.
We observe resonant reflection (Fano resonances) as well as resonant
transmission of light by optical solitons. All resonant effects can be
controlled in experiment by changing the soliton intensity.
\end{abstract}

\pacs{42.65.Tg, 05.45.-a, 42.25.Bs}

\maketitle

Recently the problem of plane wave scattering by various time-periodic potentials
has attracted much attention, as it has been shown that some
interesting effects such as resonant reflection
of waves
\cite{Cretegny} can be observed when dealing with non-stationary
scattering centers.
These resonant reflection effects were demonstrated to be similar to the well-known Fano resonance
\cite{Fano} for some nonlinear systems \cite{Flach}.
In one-dimensional (1D)
systems they can lead to the \emph{total resonant reflection} of
waves.
The time-periodic scattering centers can originate from the presence of nonlinearity
in a spatially homogeneous system \cite{DB-REVIEWS}.
The main underlying idea of this phenomenon is that a nonlinear time-periodic
scattering potential induces several harmonics. In general, these harmonics can be either inside or
outside the plane wave spectrum, 'open' and 'closed' channels, respectively \cite{Flach}.
The presence of closed channels is equivalent to a local
increase of the spatial dimensionality, i.e. to the appearance of
alternative paths for the plane wave to propagate.
This can lead to novel interference effects, such as the resonant reflection of waves - Fano resonance in nonlinear systems.

Here we emphasize, that the total resonant reflection of plane waves can
be similarly arranged by means of a static scattering center, provided the system dimensionality
has been
artificially locally increased (e.g. in composite materials)
\cite{Fano_stac}. However, such static configurations have at least two significant disadvantages.
Firstly, they do not provide
any flexibility in tuning the resonance parameters: the resonant values
of plane wave parameters are fixed by the specific geometry.
Secondly, it may be a nontrivial technological task to
introduce locally additional degrees of freedom for plane waves in the otherwise 1D system.
Time-periodic scattering potentials appear to be much more promising: they can be relatively easily
generated
(e.g. laser beams, microwave radiation, localized soliton-like excitations),
and they provide us with an opportunity to tune the resonance, since all the
resonant parameters are depending on the parameters of the
potential (e.g. amplitude, frequency) and thus can be
'dynamically' controlled by some parameter, e.g. the laser beam intensity.
\begin{figure}
\includegraphics[angle=270, width=0.45\textwidth]{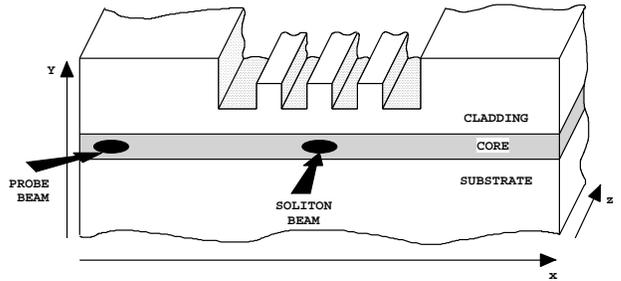}
\caption{Schematic structure of the experimental setup.
The soliton beam is sent along the $z$-axis, while the 'probe'
beam propagates in the $xz$-plane at some angle to the soliton.}
\label{fig:waveguide}
\end{figure}

In this Letter we demonstrate a possibility of experimental observation of Fano resonances in the scattering process of light by optical spatial solitons.
The soliton is generated in a planar (slab) waveguide by a laser beam injected
into the slab along the $z$-direction, see Fig.~\ref{fig:waveguide}.
The soliton beam light is confined in
the $y$-direction (inside the core layer) by the total internal reflection.
The localization
of light in the $x$-direction (the spatial soliton propagates along the $z$-direction)
is ensured by the balance between linear diffraction
and an instantaneous Kerr-type nonlinearity.
The probe beam is sent at some angle to the soliton. It has small enough amplitude
so that the Kerr nonlinearity of the medium
is negligible outside the soliton core.
It is important to note, that the process of light scattering by a spatial soliton is
assumed to be \emph{stationary} in time, i.e. the light is quasi-monochromatic.
The analogy with the above time-periodic scattering problems comes from
the possibility to interpret the spatial propagation along the $z$-direction
as an artificial time \cite{Zaharov, Agrawal}.
Thus the angle, at which the probe beam
is sent to $z$-axis plays the role of a parameter similar to
the frequency (or wave number) of plane waves in 1D systems.

In addition we allow the refractive index $n$ inside the slab
to be a stepwise function of the coordinate $x$
in a finite region, where the soliton is located,
and $n(x)\equiv n_0$ elsewhere.
The modulation of the refractive index can be
achieved by means of well-established techniques of producing waveguide arrays:
either by etching the surface of the waveguide \cite{Eisenberg},
or by inserting stripes of another
material into the slab \cite{Shimshon}.

Assuming that
the electromagnetic field maintains its linear polarization, the stationary Maxwell equation for the
Fourier component of the
electric  field
$E(x,z)$ \cite{footnote} takes the following form \cite{Ablowitz_Musslimani}:
\begin{equation}
\label{Maxwell}
\frac{\partial^2 E}{\partial z^2}+ \frac{\partial^2 E}{\partial x^2} + n^2(x) E+
\alpha \left|E\right|^2 E =0,
\end{equation}
where $\alpha>0$ is the nonlinear Kerr coefficient, the electromagnetic field is considered
to be monochromatic $\mathcal{E}(x,z,t)=E(x,z)\exp(-i\omega t)+c.c.$, and the dimensionless spatial
coordinates are used: $\omega/c = 1$, with $c$ being speed of light in vacuum.

A spatial optical soliton represents a special class of solutions of Eq.(\ref{Maxwell}) of the form
$E^{sol}(x,z)=C(x)\exp(i\beta z)$ with $C(x)$ describing the exponentially localized
profile of the soliton, $C(x)|_{x\rightarrow \pm \infty}\rightarrow 0$. $\beta$ is the only soliton
parameter ($z$-component of its wave-vector), $\beta>n_0 $. The soliton envelope function
$C(x)$ satisfies the
stationary nonlinear Schr\"odinger equation (NLS):
\begin{equation}
\label{shrod}
\frac{d^2 C}{d x^2}
+ \left[n^2(x)-\beta^2\right] C+
\alpha \left|C\right|^2 C =0,
\end{equation}

The  probe beam light interacts with the optical soliton, so that the total
electric field is a sum of two contributions, $E(x,z)=L(x,z)+C(x)\exp(i\beta z)$,
both ideally having the same frequency $\omega$.
Treating the probe beam part $L$ as a small perturbation
to the soliton part $C$ allows for a linearization of Eq.(\ref{Maxwell}) around the soliton solution,
so that we obtain:
\begin{equation}
\label{l+OB}
\frac{\partial^2 L}{\partial z^2}+ \frac{\partial^2 L}{\partial x^2}
+n^2(x)
L+\alpha\left[2|C|^2L+\mu L^*C^2e^{i2\beta
z}\right]=0.
\end{equation}
The soliton acts as an external potential for the linear beam, consisting of a 'DC' ($\sim |C|^2$) and
'AC' ($\sim C^2$) parts. While the former is the soliton induced nonlinear refractive index
modulation, the latter is the result of a
{\it coherent} interaction between the beams. Thus, the ratio of AC/DC
components depends on the {\it coherence level} between the soliton and the
probe beam, which can be controlled in experiment e.g. by a slight beam
frequency detuning or by changing the spatial correlation length
between them. The effective parameter $\mu$ ($0\le\mu\le 1$) accounts
for this possible decoherence between the beams.

The soliton part is negligible, $C(x)\rightarrow 0$,
outside the scattering center
and the refractive index is constant $n(x)\equiv n_0$. Then
Eq.(\ref{l+OB}) reduces to the simple wave equation for plane waves
$L(x,z)\sim \exp(ik_x x + i k_z z)$ with the dispersion
relation:
\begin{equation}
\label{spectr}
k_x^2+k_z^2=n_0^2.
\end{equation}

Both the modulation of the refractive index and the soliton
(the DC part of the corresponding potential to be more precise)
can result in bound states for
the probe light with the wave vector components $(k_x,k_z)$ lying outside the
spectrum (\ref{spectr}). Besides, the soliton acts as a harmonic generator (via its
AC part), so that
the general solution of the eq.(\ref{l+OB}) consists of two "harmonics" (two channels):
\begin{equation}
\label{2-channel}
L(x,z)=A(x)\exp[ik_z z]+B(x)\exp[i(2\beta-k_z)z],
\end{equation}
coupled to each other via the AC part of the soliton scattering potential
\cite{Flach}:
\begin{eqnarray}
\label{2ch-1}
A^{\prime\prime}=\left[k_z^2-n^2(x)\right]A-\alpha\left[
2|C|^2 A+\mu B^* C^2
\right],\\
B^{\prime\prime}=\left[(2\beta-k_z)^2-n^2(x)\right]B-
\alpha\left[
2|C|^2 B+\mu A^* C^2 \right].
\label{2ch-2}
\end{eqnarray}
For zero coherence parameter, $\mu=0$, the equation for  $A$ ($B$) channel is
equivalent to
the stationary Schr\"odinger equation for an effective particle
with the energy $E_A=n_0^2-k_z^2$ ($E_B=n_0^2-(2\beta-k_z)^2$) in the
external potential
being the sum of 'geometrical' and 'soliton' parts: $V_{eff}(x)=[n_0^2-n^2(x)]-
2\alpha|C(x)|^2$.

We chose $0 \le k_z \le n_0$, so that it satisfies the dispersion
relation (\ref{spectr}) with a {\em real} $k_x$ value
and the probe beam wave can freely propagate far away
from the scattering center.
Thus
the $z$-component of the wave vector in the second term in (\ref{2-channel}) $2\beta-k_z$
is outside the spectrum
(\ref{spectr}). This term corresponds
to a closed channel with amplitude exponentially decaying with the increasing distance from the scattering
center. Hence, we have one open ($A$) and one closed ($B$) channel in our scattering problem.

There is a limiting case when we can predict the position of a Fano resonance.
It corresponds
to a small coupling between the two channels provided by the AC scattering potential.
Then the Fano resonance is located at the resonance between
the discrete part of the closed channel spectrum (localized states) and the continuum part of
the open channel spectrum (delocalized states).
We take advantage of this limit by choosing
the coherence parameter
$\mu$ to be small, $\mu \rightarrow 0$ (incoherent interaction).
We will use this limit as the starting
point to catch the resonance and to follow it while increasing the AC potential strength
to its proper value for the coherent interaction, $\mu \rightarrow 1$.
By subsequently increasing
the coupling between the open and closed channels, we strongly affect the position of
localized levels in spectra of these channels. {\em Therefore even a weak modulation of the
refractive index $n$, which also affects the positions of localized levels, might
play an important role in the process of light scattering by optical solitons.}

To compute the transmission coefficient $T(k_x)$ we solve eqs.
(\ref{2ch-1},\ref{2ch-2})
with the boundary
conditions \cite{Flach}:
$A(x\rightarrow +\infty)=\tau \exp(ik_x x)$, $A(x\rightarrow -\infty)=\exp(ik_x
x) + \rho \exp(-ik_x x)$,
$B(x\rightarrow +\infty)=F\exp(-\kappa x)$, $B(x\rightarrow
-\infty)=D\exp(\kappa x)$.
Amplitudes of transmitted, $\tau$, and reflected, $\rho$, waves in the open channel
define the transmission and reflection coefficients:
$T=|\tau|^2=1-|\rho|^2=1-R$, respectively.
Amplitudes $F$ and $D$ describe spatially exponentially decaying closed channel
excitations
with the inverse localization length
$\kappa=[4\beta^2-4\beta\sqrt{n_0^2-k_x^2}-k_x^2]^{0.5}$.

\begin{figure}
\includegraphics[angle=270, width=0.4\textwidth]{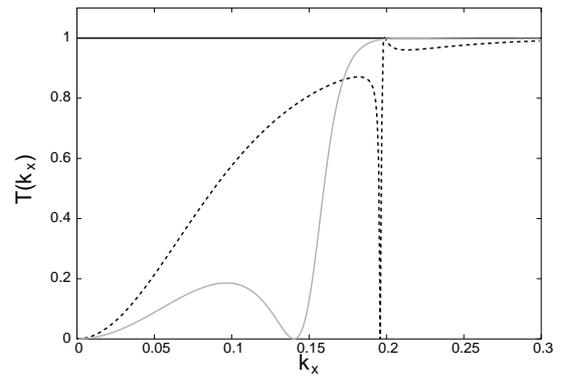}
\caption{Transmission
coefficient $T(k_x)$ of the probe beam through the optical soliton with $\beta\approx 1.4743$,
$\beta^2-n_0^2=0.1$,
in the case of a homogeneous core of the planar waveguide with $n_0=1.44$ (solid black line).
The dashed and solid gray lines indicate the transmission coefficient for the same system but for
partially decoherent beams with $\mu=0.1$ and $\mu=0.5$, respectively.}
\label{fig:transm_pure}
\end{figure}

In the simplest case of a homogeneous planar waveguide
core $n(x)\equiv n_0=const.$ $C(x)$ is the well-known stationary NLS soliton \cite{Zaharov}:
\begin{equation}
\label{nls_soliton}
C(x)=\sqrt{\frac{2\left(\beta^2-n_0^2\right)}{\alpha}}\frac{1}{\cosh\left[
\sqrt{\beta^2-n_0^2}x\right]}
\end{equation}
All resonance effects are suppressed {\it in the fully coherent case}
($\mu=1$) due to the transparency of the NLS soliton
\cite{Zaharov}, see Fig.~\ref{fig:transm_pure}, even though the resonance condition
described above can be formally satisfied for the soliton parameter $\beta$ lying in the
interval $1<\beta/n_0\lessapprox 1.5$. However, at small $\mu$ we
observe
the Fano resonance at wavenumbers $k_x$ close to the predicted position in the limit of
a small interchannel coupling (dependent on the soliton parameter $\beta$), see Fig.~\ref{fig:transm_pure}.
Increasing coherency $\mu$ leads to a shift of the resonance position towards the
band edge $k_x=0$ (this limiting
value corresponds to the 'probe' beam being sent along the $z$-axis, i.e.
parallel to the soliton). Finally, when
the soliton and the probe beam are coherent ($\mu=1$), we loose the resonance
completely and the soliton becomes transparent ($T\equiv 1$).

In order to observe Fano resonances for coherent beams we consider a
non-homogeneous refractive index $n(x)$. Introducing modulation of the
refractive index, the soliton is no longer transparent for the probe beam.
However, this does not automatically guarantee appearance of resonances,
thus specific core configurations must be designed.

\begin{figure}
\includegraphics[angle=270, width=0.4\textwidth]{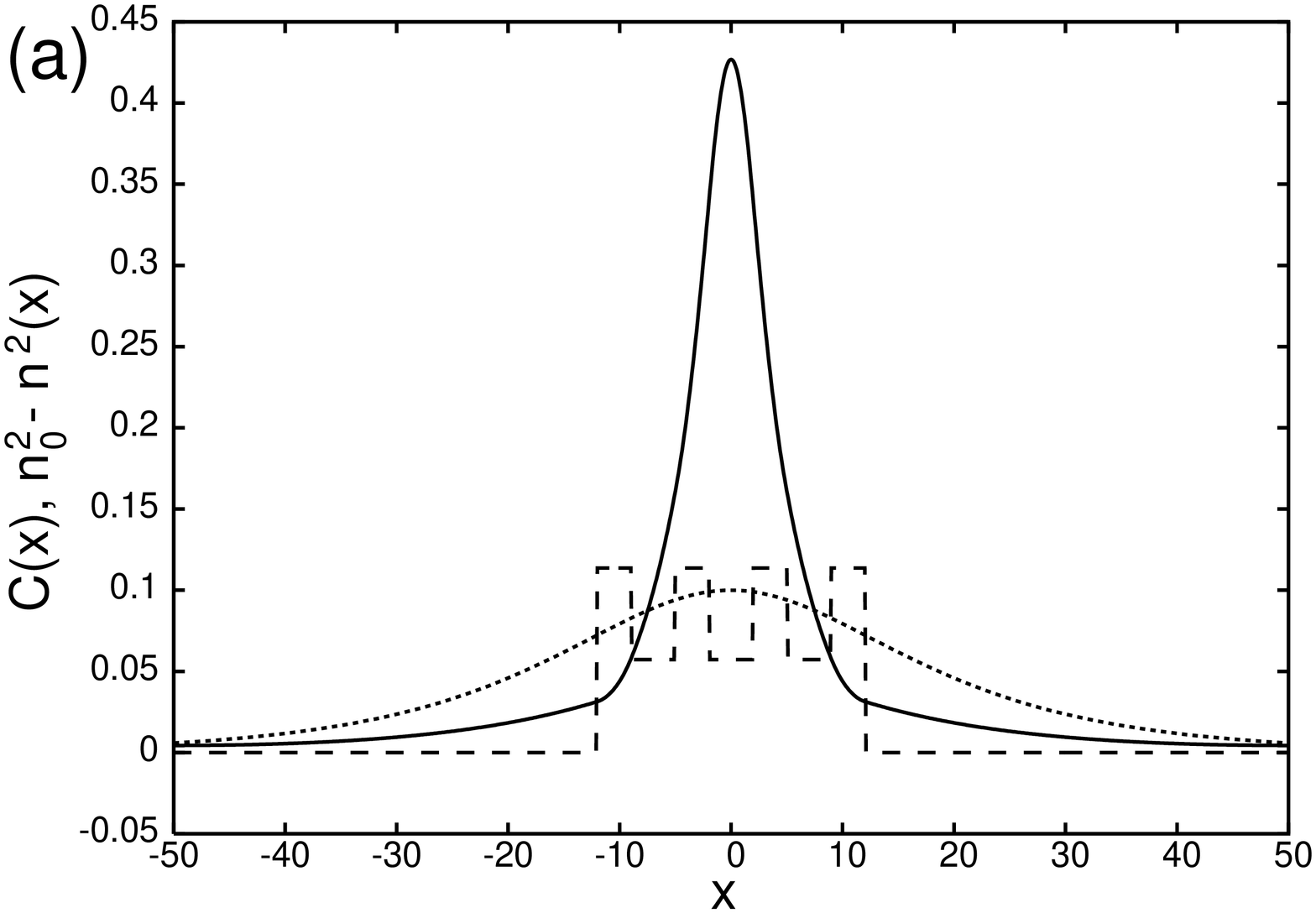}
\includegraphics[angle=270, width=0.4\textwidth]{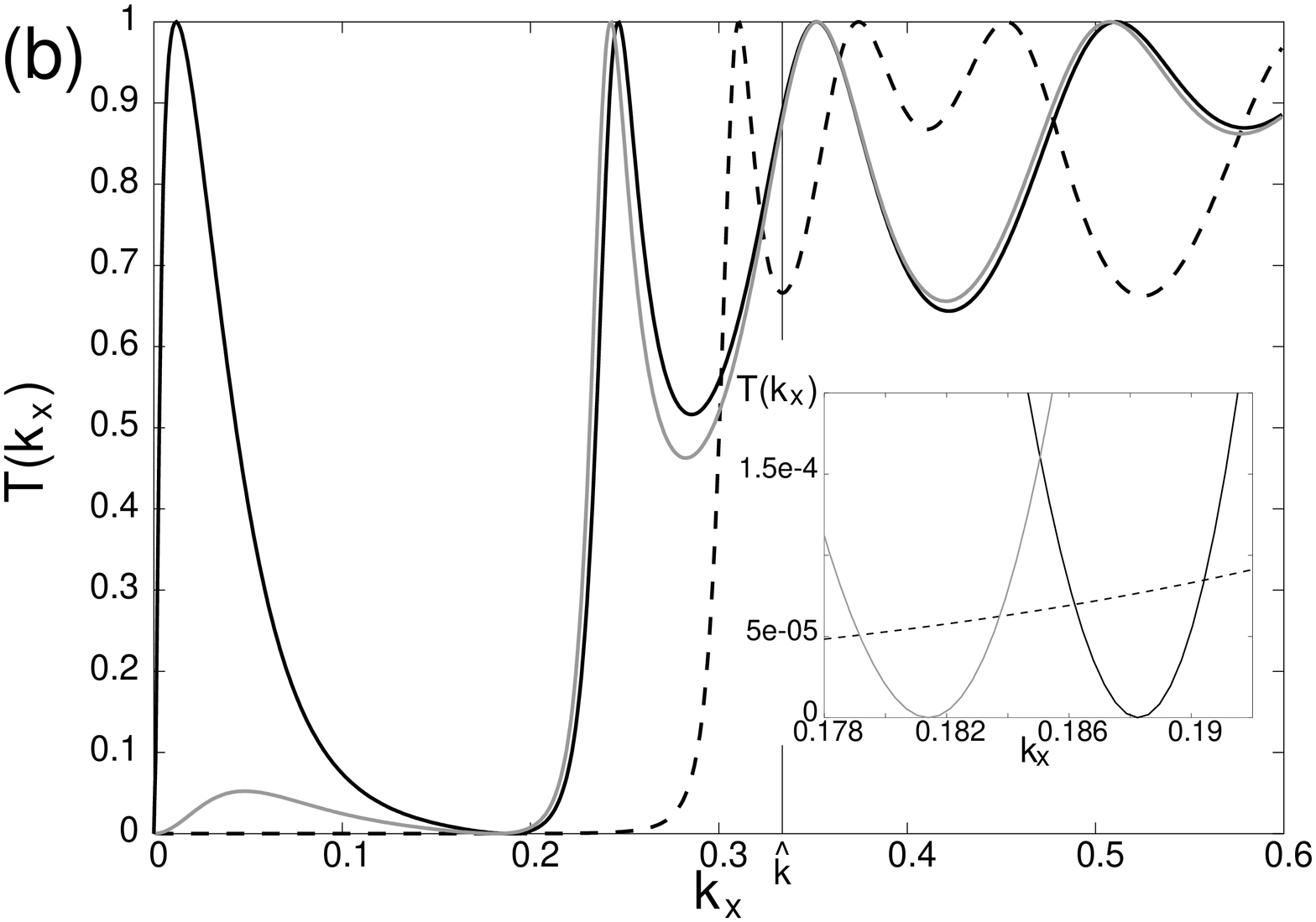}
\caption{(a) The effective potential $n_0^2-n^2(x)$ (dashed line) and the soliton
profile $C(x)$ (solid line) for the
TW configuration,
parameter values are: $n_0=1.44$, $n_1=1.4$, $n_2=1.42$, $L=12$,
$L_b=4$, $\beta \approx 1.4417$, $\beta^2-n_0^2=5.0e$-3.
The dotted line shows the soliton profile for
the case of a homogeneous core $n(x)\equiv n_0=1.44$ with the same value of the soliton
parameter $\beta$; (b) Transmission
coefficient $T(k_x)$ for the system without a soliton (dashed line), and with solitons
having different parameter values:
$\beta\approx 1.4417$, $\beta^2-n_0^2=5.0e$-3 (solid black line), $\beta \approx 1.4434$,
$\beta^2-n_0^2=1.0e$-2 (solid gray line). Vertical line indicates the critical value
$\hat{k}\approx 0.337$
(see the main text for details).
Note, that $\mu=1$ here (coherent interaction).}
\label{fig:triplet}
\end{figure}

We propose here a triple well (TW) configuration of the planar waveguide core,
which allows one to observe the Fano resonance effect in light scattering by an optical soliton.
The refractive index $n(x)$ inside the core is locally decreased in the vicinity of the scattering
center (i.e. where an optical soliton is formed), $n(x)=n_1<n_0, |x|<L$. In addition, to
stabilize the soliton, we introduce
local 'wells' with a higher value of the refractive index $n=n_2$
($n_1<n_2<n_0$) inside the $n=n_1$ section, the width of each well is $L_b<L$.
It would suffice to insert only one well, but the TW configuration provides one
with more pronounced Fano resonances.
The resulting structure
of the effective potential $n_0^2-n^2(x)$, caused by the refractive index modulation,
and the stable optical soliton profile $C(x)$ are shown
in Fig.~\ref{fig:triplet}(a).
The transmission coefficient computed for the system both with and without solitons with different parameters is plotted in Fig.~\ref{fig:triplet}(b).

Note, that the $T(k_x)$ dependence for the TW configuration without a soliton
has several peculiarities. First of all, there is always a critical value $\hat{k}$, below
which the transmission coefficient is rather small. This has a direct connection to the well
known effect of the 'total internal reflection' at the boundary between the
homogeneous core and
the central section with a \emph{lower} value of the refractive index. In addition,
we
observe several transmission peaks at larger
$k_x$ values connected with internal modes and
determined by the specific internal configuration of the TW section
[i.e. by the $n(x)$ dependence] \cite{Sprung}.

Switching on a spatial optical soliton has a twofold effect on the $T(k_x)$ dependence.
We add both the DC part and coupling
to the closed channel mediated by the AC part of the corresponding effective
potential in (\ref{l+OB}) to the scattering potential.
The Kerr nonlinearity changes
the effective refractive index in the scattering center region:
$n^2_{eff}(x)=n^2(x)+\alpha|C(x)|^2$. This, in turn,
leads to a total restructuring of all the internal modes, and therefore all
the resonant transmission peaks are shifted, as seen
in Fig.~\ref{fig:triplet}(b). This effect has a 'stationary' nature, meaning that
only the soliton profile $C(x)$ rather than its phase is responsible for the new positions
of the resonance transmission peaks.

Most important, one can clearly observe
appearance of the resonant reflection ($T=0$) of the probe
beam in the 'dark region' $k_x<\hat{k}$
[see inset in Fig.~\ref{fig:triplet}(b)]. Thus, the specially designed configuration of
the planar waveguide core allows one
to observe Fano resonances for {\it coherent scattering}, provided that the
soliton intensity (which is directly connected to its parameter $\beta$) is not too high.
At low soliton intensities (small values of $\beta$) we also observe an
additional resonance transmission peak ($T=1$) close to the position of the Fano resonance
[solid black curve in Fig.~\ref{fig:triplet}(b)]. Such a perfect transmission resonance generally
accompanies Fano resonances when the coupling between the open and closed channels is small \cite{Flach}.
As the soliton intensity increases,
and therefore the strength of coupling between the open and closed channels also increases,
the Fano resonance position, together with the resonance transmission position, are shifted
towards the lower band edge ($k_x=0$). For higher values of the soliton intensity only resonant
reflection is observed, while the resonance transmission peak is already outside of the allowed
wavenumber range,
see solid gray curve in Fig.~\ref{fig:triplet}(b). Finally, at high enough
soliton intensities both resonance effects disappear, so that for $\beta \approx 1.47$ ($\beta^2-n_0^2=0.1$)
we observe only the above 'stationary' effect of shifting of all the
resonance transmission peaks connected to the internal modes.

To conclude, we propose experimental setups for a direct observation
of Fano resonances in scattering of light by optical solitons.
Such experiments would be
of a great importance, as they could confirm several theoretical predictions of
resonance phenomena in plane wave scattering by localized nonlinear excitations
\cite{Flach}.
The scattering
process is performed in a planar waveguide core. Two possible ways to obtain
resonances in the transmission are either to introduce a
decoherence between the probe and the soliton beams, or to insert a
specially designed non-homogeneous refractive index section inside the core.
In both cases all the resonance effects, predicted by our analysis, can be easily
tuned in experiment,
as they strongly depend on the soliton intensity. This could be also of importance
from the practical point of view, giving an opportunity to use such resonance effects in
different optical switchers for high-speed optical communication devices and also in extensively
developing area of optical logic devices.

\begin{acknowledgments}
We would like to thank Shimshon Bar-Ad and Mordechai Segev for useful and
stimulating discussions.
\end{acknowledgments}


\begin{thebibliography}{99}
\bibitem{footnote}We consider the $y$-dependence of the field to be
quasi-uniform all over the system. Weak deviations from the uniform distribution,
caused by a modulation of the
refractive index $n$, are not essential, since they don't play any significant role
in the resonant effects we describe. A stepwise change of the
refractive index leads to an excitation of radiative modes in the vicinity, where this change takes place. The associated rate of losses is proportional to the refractive index variation. In our case it is of the order of 1\%.
\end{thebibliography}
\end{document}